\documentclass[runningheads]{llncs}
\usepackage{graphicx}
\usepackage{amsmath,amssymb}
\usepackage{mathrsfs}
\usepackage{amsmath}
\begin{document}
\title{Micro CT Image-Assisted Cross Modality Super-Resolution of Clinical CT Images Utilizing Synthesized Training Dataset}
%
%
\author{Tong Zheng\inst{1}, \and
Hirohisa Oda\inst{2}, Shota Nakamura\inst{2}, Masahiro Oda\inst{1},
\\ Kensaku Mori \inst{1}}
%
%
\authorrunning{***** et al.}
\titlerunning{Micro CT Image-Assisted Cross Modality Super-Resolution}
\institute{Graduate School of Informatics, Nagoya University, Japan \and
Nagoya University Graduate School of Medicine, Japan}

\maketitle              
\begin{abstract}
This paper proposes a novel, unsupervised super-resolution (SR) approach for performing the SR of a clinical CT into the resolution level of a micro CT ($\mu$CT). The precise non-invasive diagnosis of lung cancer typically utilizes clinical CT data. Due to the resolution limitations of clinical CT (about 0.5$\times$0.5$\times$0.5 mm$^3$), it is difficult to obtain enough pathological information such as the invasion area at alveoli level. On the other hand, $\mu$CT scanning allows the acquisition of volumes of lung specimens with much higher resolution (50$\times$50$\times$50 $\mu$m$^3$ or higher). Thus, super-resolution of clinical CT volume may be helpful for diagnosis of lung cancer. Typical SR methods require aligned pairs of low-resolution (LR) and high-resolution (HR) images for training. Unfortunately, obtaining paired clinical CT and $\mu$CT volumes of human lung tissues is infeasible. Unsupervised SR methods are required that do not need paired LR and HR images. In this paper, we create corresponding clinical CT - $\mu$CT pairs by simulating clinical CT images from $\mu$CT images by modified CycleGAN. After this, we use simulated clinical CT - $\mu$CT image pairs to train an SR network based on SRGAN. Finally, we use the trained SR network to perform SR of the clinical CT images. We compare our proposed method with another unsupervised SR method for clinical CT images named SR-CycleGAN. Experimental results demonstrate that the proposed method can successfully perform SR of clinical CT images of lung cancer patients with $\mu$CT level resolution, and quantitatively and qualitatively outperformed conventional method (SR-CycleGAN), improving the SSIM (structure similarity) form 0.40 to 0.51.

\end{abstract}
\section{Introduction}
Lung cancer is now the most common cancer among men worldwide \cite{lungcancer}. Precise non-invasive diagnosis of lung cancer mainly relies on clinical CT images \cite{silvestri2013methods}. However, due to the resolution limitations of clinical CT (about 0.5$\times$0.5$\times$0.5 mm$^3$), it is difficult to obtain enough pathological information such as the invasion area at alveoli level. $\mu$CT volumes obtained by $\mu$CT scanning of resected lung cancer specimens can capture detailed and surrounding anatomical structures of them. For more precise clinical CT diagnosis including diagnosing the areas invaded by cancer, super-resolution (SR) of clinical CT image into $\mu$CT level would be one of the options. However, most SR methods require paired training datasets (in form of spatially registrated clinical and $\mu$CT volumes) which are not feasible to collect.  

Typical SR methods are regarded as supervised, which require aligned pairs of LR and HR images for training \cite{dong2015image,aggarwal2018modl,johnson2016perceptual,ledig2017photo}. There are only a few unsupervised SR methods that do not require paired LR and HR images \cite{ravi2019adversarial,zheng2019multi,lugmayr2019unsupervised,yuan2018unsupervised}. 
Ravia et al. \cite{ravi2019adversarial} proposed an unsupervised image SR method for endomicroscopy.  However, it requires the fiber positions in endoscope's cable as additional input; since CT and $\mu$CT  images are acquired with different devices (CT scanners), it is not possible to adapt this approach. Zheng et al. proposed an an unsupervised clinical CT image SR method called SR-CycleGAN \cite{zheng2019multi}. However, this method is shown to be difficult to train and can produce severe noise in the SR results. Lugmayr et al. \cite{lugmayr2019unsupervised} proposed a unsupervised method for real-world super-resolution. However, this approach can only perform SR of images form one domain to another similar domain (e.g. low-resolution images to original HR images). Since $\mu$CT images and clinical CT images are in totally different domains (shot with different devices), we consider Lugmayr's method needs to be modified an adopted for medical image super-resolution.

In this paper, we address the problem where there is no paired clinical and $\mu$CT dataset by generating synthesized paired training datasets. First, we create synthesized clinical CT images from $\mu$CT images utilizing a modified CycleGAN \cite{zhu2017unpaired} approach with additional SSIM \cite{wang2004image}-based loss term. Note that these synthesized clinical CT images are paired with original $\mu$CT images. Subsequently, we utilize paired $\mu$CT - synthesized clinical CT images for training a supervised SR network. As reference, we use the trained SR network for performing SR of the clinical CT images.

The following are the the contributions of this paper: 1) trans-modality super-resolution from clinical CT to $\mu$CT-level and 2) an SR approach for clinical CT that can be operated without the need of any paired LR-HR data.

\section{Methods}
\subsection{Overview}
Our proposed method performs SR of clinical CT images into $\mu$CT scale. This process consists of two parts. The first part is a trans-modality network (synthesize network) which transforms micro-CT images into clinical CT-like images for building synthesized clinical CT - $\mu$CT dataset. The second part is a super-resolution network (SR network) which learns a mapping from clinical CT-like images (LR images) to micro CT images (HR images), trained on synthesized clinical CT - $\mu$CT dataset. First we adapt a preprocessing (explained below) to the clincal CT and $\mu$CT data. After the preprocessing, a $\mu$CT image $\boldsymbol{x}$ is the input of the synthesize network. Modality-translation is applied to $\boldsymbol{x}$ by synthesize network to generate clinical CT-like images $\boldsymbol{\hat{y}}$. $\mu$CT images $\boldsymbol{x}$ and clinical CT-like images $\boldsymbol{\hat{y}}$ are used to create synthesized clinical CT - $\mu$CT dataset for training the SR network. For reference, we only use the SR network. A clinical CT image of $i \times i$ (pixels) is the input of the SR network. The output is an SR image of $j \times j$ pixels ($j=8i$). At last we adapt a postprocessing to the SR image.

For the preprocessing, we use region growing to extract lung area form clinical CT volumes. We normalize intensity range of clinical CT and $\mu$CT images to -1 and +1. We randomly crop 2D image patches whose sizes are $i \times i$ (pixels) from clinical CT volumes and $j \times j$ from $\mu$CT volumes ($j=8i$), for training of the network. For the postprocessing, we jointly combined output SR images of $j \times j$ (pixels) to reconstruct the whole SR CT image.

\subsection{Synthesize Clinical CT-like Images from $\mu$CT Images}
\subsubsection{Synthesize Network}
Our goal is to perform SR of clinical CT into $\mu$CT level. However, we cannot obtain any paired clinical - $\mu$CT images since registration between clinical CT and $\mu$CT is very challenging. As an alternative, we consider synthesizing clinical CT-like images from $\mu$CT images, as to obtain synthesized clinical - $\mu$CT pairs. We aim to learn a mapping $G_1(X)$ that maps images from $\mu$CT domain $X$ to clinical CT domain $Y$. Note that the clinical CT images cannot be synthesized by directly downsampling $\mu$CT images since they are acquired with different devices. 

We design the mapping $G_1(X)$ from the $\mu$CT domain to the clinical CT domain based on a FCN. Following CycleGAN \cite{zhu2017unpaired}, we define a network $G_1$ as a mapping that maps an image from downsampled $\mu$CT images domain $f(X)$ to clinical CT image domain $Y$; a network $G_2$ maps images from clinical CT image domain $Y$ to the downsampled $\mu$CT image domain $f(X)$. Here $f()$ is defined as a gaussian pyramid downsampling function. We define a discriminator network $D_1$ distinguishs synthesized clinical CT-like images $\hat{\boldsymbol{y}} = G_1(f(\boldsymbol{x}))$ and real clinical CT images $\boldsymbol{y}$; a discriminator network $D_2$ to distinguish $\mu$CT-like images $f(\hat{\boldsymbol{x}})=G_2(\boldsymbol{y})$ and real downsampled $\mu$CT images $f(\boldsymbol{x})$ from $\mu$CT images $\boldsymbol{x}$.

Compared with the conventional CycleGAN approach, we add SSIM (structure similarity) \cite{wang2004image} as an additional loss term. We define the SSIM loss between images $\boldsymbol{a}$ and $\boldsymbol{b}$ as:
\begin{equation}
\mathcal{L}_{\rm S}(\boldsymbol{a},\boldsymbol{b}) =  1-\frac{(\mu_{\boldsymbol{a}}\mu_{\boldsymbol{b}}+C_1)(2\sigma_{\boldsymbol{ab}}+C_2)}{(\mu_{\boldsymbol{a}}^2+\mu_{\boldsymbol{b}}^2+C_1)(\sigma_{\boldsymbol{a}}^2+\sigma_{\boldsymbol{b}}^2+C_2)},
\end{equation}
here $\mu_{\boldsymbol{a}}$ and $\mu_{\boldsymbol{b}}$ denotes the average intensity of image $\boldsymbol{a}$ and $\boldsymbol{b}$. $\sigma_{\boldsymbol{a}}$ and $\sigma_{\boldsymbol{b}}$ are variance of image $\boldsymbol{a}$ and $\boldsymbol{b}$ respectively. $\sigma_{\boldsymbol{ab}}$ denotes the covariance of images $\boldsymbol{a}$ and $\boldsymbol{b}$.  $C_1$ and $C_2$ are constant numbers included to avoid instability.

\begin{figure}[bt]
\begin{center}
\includegraphics[width=0.9\textwidth]{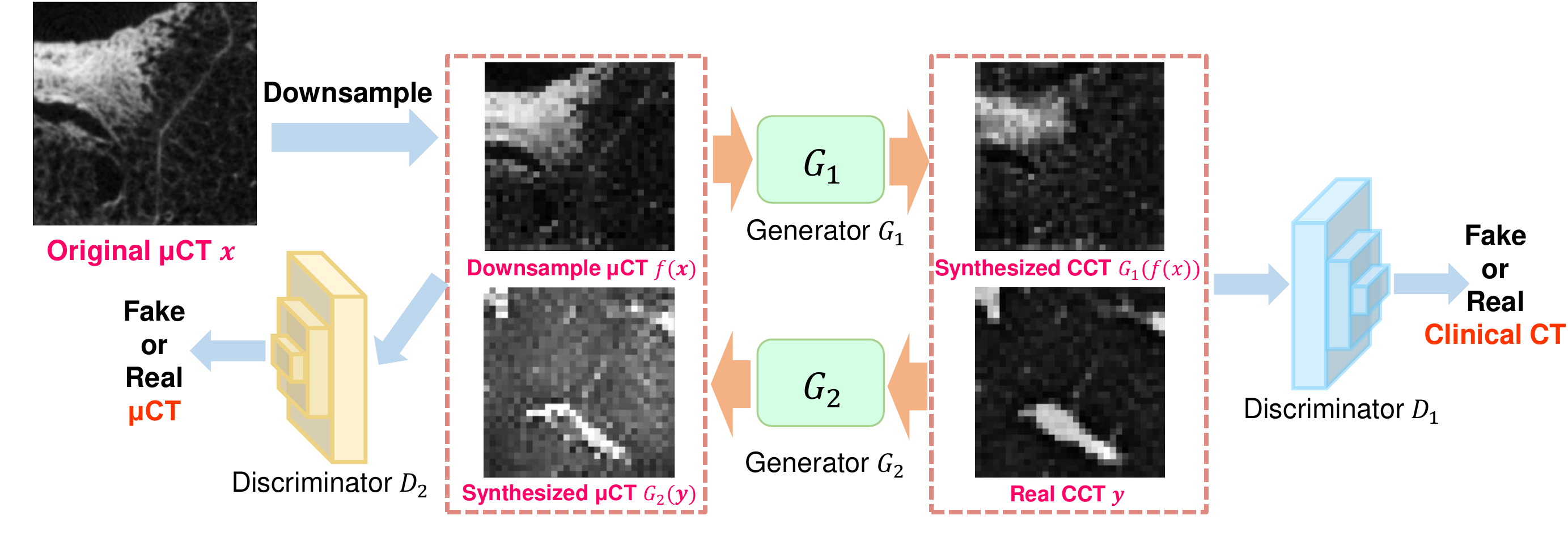}
\end{center}
\caption{Structure of synthesize network used for synthesizing clinical CT (CCT) from $\mu$CT. Proposed approach consists of the following steps: (1) downsample original $\mu$CT $\boldsymbol{x}$ to downsampled $\mu$CT $f(\boldsymbol{x})$; (2) use downsampled $\mu$CT $f(\boldsymbol{x})$ and real clinical CT patches $\boldsymbol{y}$ to train CycleGAN, including training mapping $G_1$ and $G_2$, and (3) trainned mapping network $G_1$ will be used for generating clinical CT-like images from $\mu$CT images. } \label{CycleGANfig}
\end{figure}

The overall loss function can be formulated as:
\begin{equation}
\begin{split}
    \mathcal{L}(G_1,G_2,D_1,D_2) &= \mathcal{L}_{O}(G_2,G_2,D_1,D_2) \\ 
                                 &+ \lambda_1 \mathcal{L}_S(f(\boldsymbol{x}),G_1(f(\boldsymbol{x}))) + \lambda_2 \mathcal{L}_S(\boldsymbol{y}, G_2(\boldsymbol{y})),
\end{split}
\end{equation}
here $\mathcal{L}_{O}$ is the loss function of the original CycleGAN, and $\mathcal{L}_S(f(\boldsymbol{x}),G_1(f(\boldsymbol{x})))$ and $\mathcal{L}_S(\boldsymbol{y}, G_2(\boldsymbol{y}))$ are SSIM loss terms. The structure of proposed synthesize network is shown in Fig.~\ref{CycleGANfig}. 

\subsubsection{Network Training}
We train the synthesize network using clincial CT and $\mu$CT images. Downsampled $\mu$CT images $f(\boldsymbol{x})$ ($f()$ is the downsample function) are are fed to the network $G_1$. The synthesized clinical CT images $\hat{\boldsymbol{y}}$ which are correspond to $\boldsymbol{x}$ are generated. On the other hand, clinical CT images $\boldsymbol{y}$ are fed to network $G_2$. The synthesized micro CT images $f(\hat{\boldsymbol{x}})$ which are correspond to $\boldsymbol{y}$ are generated. 

\subsubsection{Building Synthesized Clinical CT - $\mu$CT Dataset}
We use trained generator $G_1$ in for synthesizing clinical CT-like images from $\mu$CT images. $\mu$CT images $\boldsymbol{x}$ are are fed to the network $G_1$ to generate corresponding synthesized Clinical CT images $\hat{\boldsymbol{y}}$. Large amount of corresponded $\boldsymbol{x}$ and $\hat{\boldsymbol{y}}$ forms synthesized clinical CT - $\mu$CT dataset, which is used for training SR network. 
\begin{figure}[bt]
\begin{center}
\includegraphics[width=0.9\textwidth]{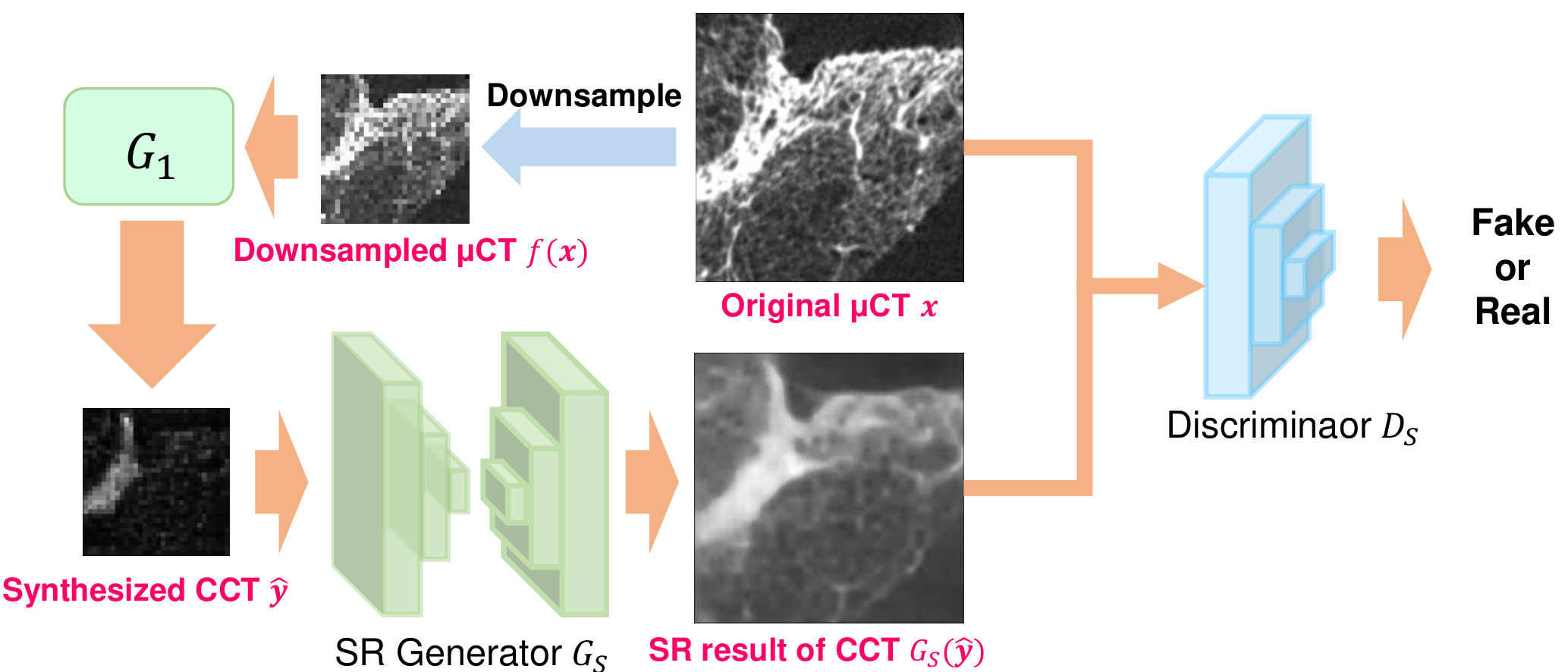}
\end{center}
\caption{Structure of our proposed SR network. First, we use gaussian pyramid downsample function $f()$ to downsample original $\mu$CT $\boldsymbol{x}$ to 1/8 of its original resolution. Next, we use the trained generator $G_1$ from synthesize network (for synthesizing clinical CT-like images from $\mu$CT images) to synthesize clinical CT-like images $\hat{\boldsymbol{y}}$ from downsampled $\mu$CT images $f(\boldsymbol{x})$. Finally, we use adversarial training to learn the SR network $G_S$ that super-resolves synthesized clinical CT $\hat{\boldsymbol{y}}$ to $G_S(\hat{\boldsymbol{y}})$ using the corresponding $\mu$CT $\boldsymbol{x}$ as ground truth. Note that while training the SR network, weights of network $G_1$ are fixed.} \label{SRGANfig}
\end{figure}

\subsection{Super-resolution of Clinical CT Images using Synthesized Training Data}
\subsubsection{SR Network}
By building synthesized clinical CT - $\mu$CT dataset, we are able to train a supervised SR network using synthesized clinical CT-like images $\hat{\boldsymbol{y}}=(G_1(f(\boldsymbol{x})))$ and corresponding $\mu$CT images $\boldsymbol{x}$. We want to learn a network that can map images $\boldsymbol{\hat{y}}$ in the synthesized clinical CT domain $\hat{Y}$ to images $\boldsymbol{x}$ in $\mu$CT domain $X$.

Following the SRGAN approach \cite{ledig2017photo}, we also use a GAN-based network to perform SR. A super-resolution network $G_S$ performs SR of the input images. A discriminator network $D_S$ differentiates SR images $G_S(\hat{\boldsymbol{y}})$ and real $\mu$CT images $\boldsymbol{x}$. The discriminator generator networks are trained alternatively to minimize two loss terms. The first loss term is the pixel-wise $l_2$ loss between a input clinical CT-like image and the desired $\mu$CT output:
\begin{equation}
\mathcal{L}_{2}(G_S) = \mathbb{E}_{\hat{\boldsymbol{y}} \sim P(\hat{\boldsymbol{y}}), \boldsymbol{x} \sim P(\boldsymbol{x})} \|G_S(\hat{\boldsymbol{y}}) - \boldsymbol{x}\|_2^2,
\end{equation}
where $P(\hat{\boldsymbol{y}})$ is the distribution of the generated clinical CT-like images, and $P(\boldsymbol{x})$ is the distribution of the micro CT images. This loss term constraints pixel-wise similarity between the input LR images and the desired output HR images. We define a second loss term that represents the adversarial loss for training the generator $G_S$:
\begin{equation}
\mathcal{L}_{A}(G_S) = \mathbb{E}_{\hat{\boldsymbol{y}} \sim P(\hat{\boldsymbol{y}})}[-\log(D_S(G_S(\hat{\boldsymbol{y}}))].
\end{equation}
This loss term constraints generator $G_S$ for generating more realistic images that are close enough to $\mu$CT domain $Y$ as to fool the discriminator $D_S$. The following is the total objective function for training SR generator $G_S$: 
\begin{equation}
\mathcal{L}(G_S) = \mathcal{L}_{2} + \lambda \mathcal{L}_{A},
\end{equation}
where $\lambda$ is the weight of adversarial loss term $\mathcal{L}_{A}$. The structure of the GAN-based SR network is shown in Fig.~\ref{SRGANfig}.

\subsubsection{Network Training}
We  train  the  synthesize  network  using synthesized clinical CT - $\mu$CT dataset. Synthesized clinical CT images $\boldsymbol{\hat{y}}$ are fed to network. Output is a SR image $G_S(\boldsymbol{\hat{y}})$.

\subsubsection{SR of clinical CT images}
To perform SR of a clinical CT image $\boldsymbol{y}$, $\boldsymbol{y}$ is input to trained SR generator $G_S$. Then we obtain output $G_S(\boldsymbol{y})$, which is the SR image.



\begin{figure}[bt]
\begin{center}
\includegraphics[width=0.99\textwidth]{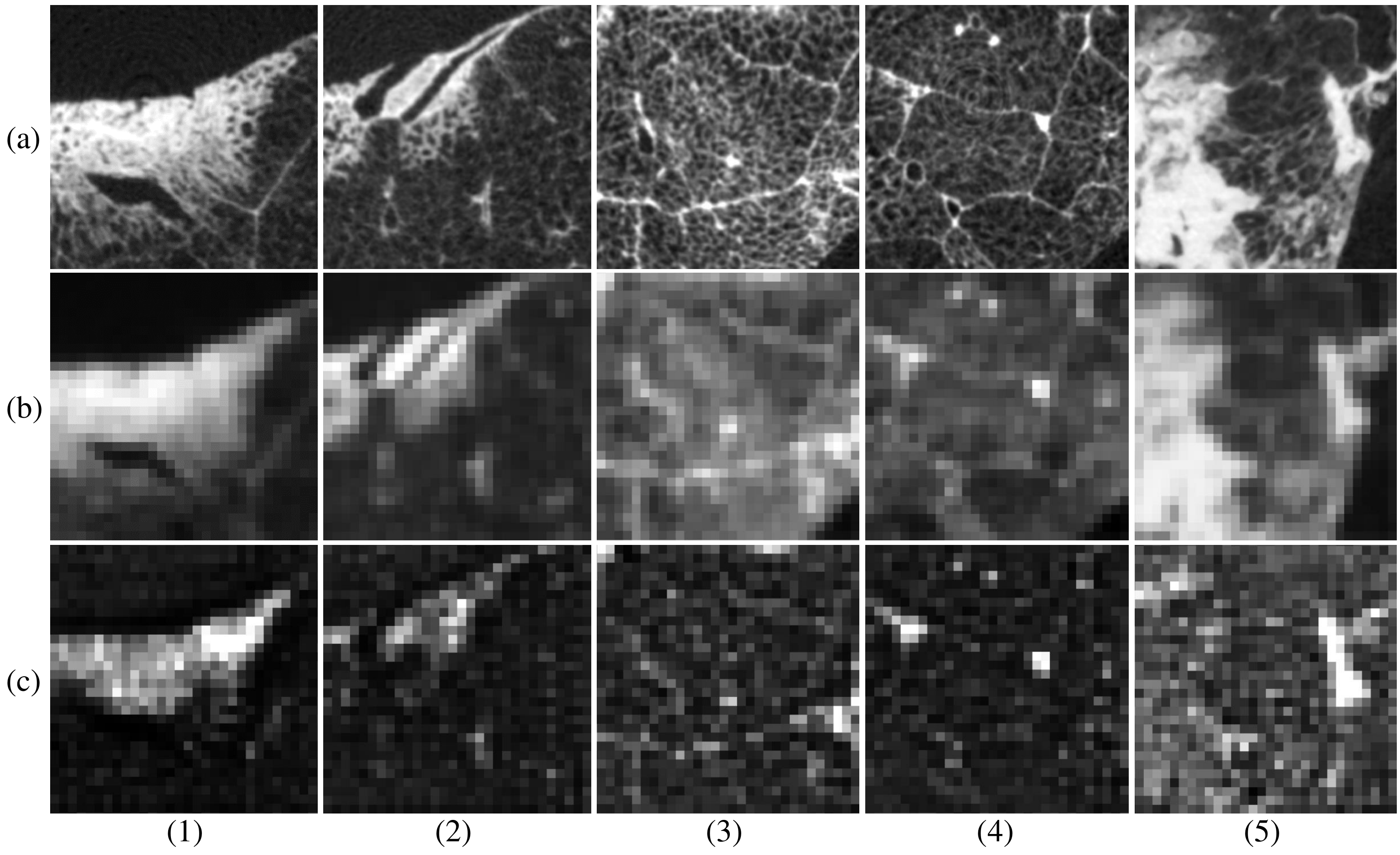}
\end{center}
\caption{Row (a): image slices cropped from $\mu$CT volumes; row (b): images downsampled from images from row (a); row (c): synthesized images generated by synthesis network's generator $G_1$ from images of row (b). Columns (1) and (2) are cropped from one case, and columns (3) and (4) are cropped from another. Column (5) is cropped from another. $\mu$CT images are translated to clinical CT style like in row (3), while image structure remains nondestructived. Images of row (c) are used as input for training the SR network using images of row (a) as corresponding ground truth.} \label{Synthsispairs.fig}
\end{figure}

\section{Experiments and Results}
For qualitative evaluation, we applied the method to clinical CT images to obtain SR images. For quantitative evaluation, we applied the method to synthesized clinical CT images to obtain SR images, then compare SR images and corresponding $\mu$CT images by SSIM.
\subsection{Dataset}
First, we utilize eight cases of clinical CT volumes and six cases of $\mu$CT volumes for training the synthesize network. Second, we utilize five clinical CT volumes and six $\mu$CT volumes for training the SR network. The $\mu$CT volumes are of cancer specimens obtained after lung resection surgeries. The clinical CT volumes are acquired using a clinical CT scanner (SOMATOM Definition Flash, Siemens Inc., Munich) with a resolution of 625$\times$625$\times$600 $\mu$m$^3$ / voxel. The $\mu$CT volumes were acquired using a $\mu$CT scanner (inspeXio SMX90CT Plus, Shimadzu, Kyoto), with isotropic resolutions in the range of 42-52 $\mu$m.

\subsection{Parameter Settings}
During training, we extract 2000 patches from each case. Based on the number of pixels in the lung area in clinical CT images, the size of patches extracted from the clinical CT volumes were of 32$\times$32 pixels. The size of patches extracted from the $\mu$CT volumes were of 256$\times$256 pixels. Since the super-resolution always enlarged the input images to power of 2 times (2, 4, 8 times, e.g.), and comparing the resolution of the the clinical CT volumes (625$\mu$m) is about ten times of the $\mu$CT volumes (52$\mu$m), we considered 8 times to be the most proper. For SSIM loss, we set $C_1$ and $C_2$ as 0.02 and 0.06, respectively. For wights of loss terms of synthesize network, we set $\lambda_1$ to 0.5 and $\lambda_2$ to 0.4. For wights of loss terms of loss SR network, we set $\lambda$ to 0.001. Epoch number of both synthesize network and SR network is 200 with a 64-minibatch size. 

\subsection{Separate Training of Synthesize Network and SR Network}
We separately train the synthesize network and the SR network. First, the synthesize network is trained to generate clinical CT-like from $\mu$CT images (Synthesized clinical CT - $\mu$CT pairs by are illustrated in Fig.~\ref{Synthsispairs.fig}.); second, the SR network is trained using synthesized clinical CT - $\mu$CT images.

\begin{figure}[bt]
\begin{center}
\includegraphics[width=0.99\textwidth]{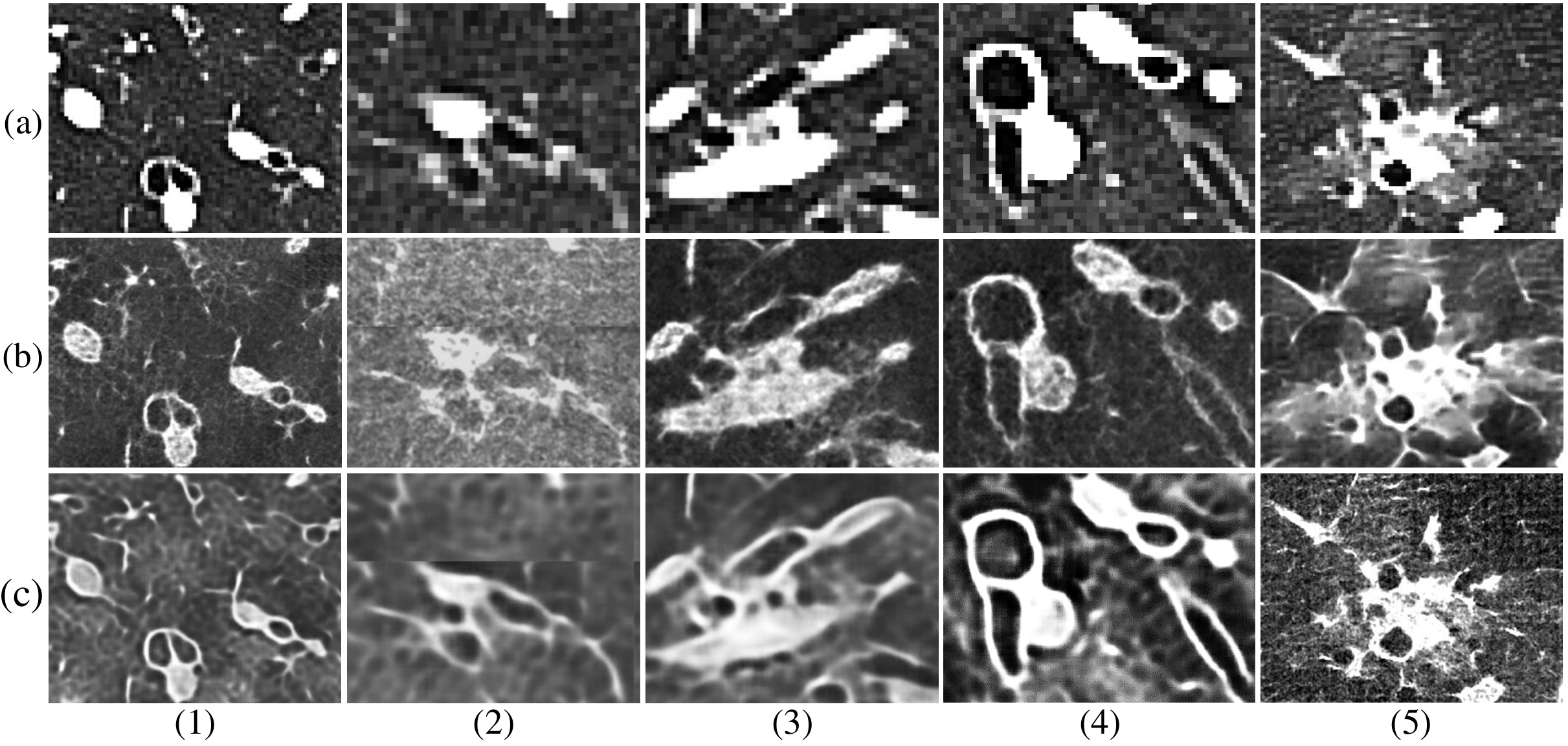}
\end{center}
\caption{Row (a): original clinical CT images; (b): SR result of conventional method (SR-CycleGAN); (c): SR result of proposed method. Columns (1), (2), (3), and (4) are images cropped from bronchus region, (5) are images cropped from tumor region.} \label{SRclinical.fig}
\end{figure}

\subsection{Results}

\begin{table} [bt]
\center
\caption{Proposed method greatly outperformed SR-CycleGAN in SSIM (a standard for evaluating similarity of two given images)}\label{tab1}
\setlength{\tabcolsep}{7mm}
\begin{tabular}{|l|l|l|}
\hline
  Method & SSIM\\
\hline
SR-CycleGAN \cite{zheng2019multi} & 0.40\\
Proposed Method & {\bfseries0.51}\\
\hline
\end{tabular}
\end{table}

\subsubsection{Quantitative Evaluation}
We used two $\mu$CT volumes for quantitative evaluation. Since we do not have any corresponding clinical CT - $\mu$CT pairs, we propose a novel quantitative evaluation method: first we use the trained generator network $G_1$ of synthesize network to generate a clinical CT-like images $G_1(f(\boldsymbol{x}))$ from a $\mu$CT images $\boldsymbol{x}$, and then use trained SR generator $G_S$ to obtain SR image $G_S(G_1(f(\boldsymbol{x})))$. We utilized SSIM \cite{wang2004image} to compare the SR image and the original $\mu$CT image of the conventional method (SR-CycleGAN) and our proposed method. Table \ref{tab1} illustrates the quantitative results of both methods.

\subsubsection{Qualitative Evaluation}
We used two clinical CT volumes for our qualitative evaluation. 
We utilize the synthesised clinical - $\mu$CT pairs for training the SR network. We use the trained SR network for performing SR of the clinical CT images. We compare the SR results of proposed method with the SR-CycleGAN approach \cite{zheng2019multi}. The results are shown in Fig.~\ref{SRclinical.fig}.

\section{Discussion and Conclusion}
To the best of our knowledge, our is the first study that performs SR of clinical CT images by training using synthesized clinical CT - $\mu$CT pairs. 
Important anatomical structures such as bronchus, vein and contour of tumor are reconstructed clearly as in Fig~\ref{SRclinical.fig}: walls of bronchioles and veins become smother, and size of tumor become clearer, compared to SR-CycleGAN's result. 
We consider this is because the proposed method consists of two networks for different jobs (one for modality transformation and another for SR). On the other hand, SR-CycleGAN combined modality transformation and SR in one single network, which causes training to be unstable and generate noisy results.

In this scheme, the only feasible quantitative evaluation approach is conducted by synthesizing clinical CT - $\mu$CT pairs and evaluates how well synthesized clinical CT can be reconstructed to $\mu$CT CT images. Identifying more convincing quantitative evaluation methods is our future work.

In this paper, we have proposed a novel unsupervised (SR) approach for performing the SR of clinical CT images. By Synthesizing clinical CT images from $\mu$CT images, we solved the problem of no paired clinical CT - $\mu$CT data. The proposed method outperformed conventional method qualitatively and quantitatively. The results demonstrates that our proposed method successfully performed SR of lung clinical CT images into $\mu$CT level.

\begin{figure}[bt]
\begin{center}
\includegraphics[width=0.99\textwidth]{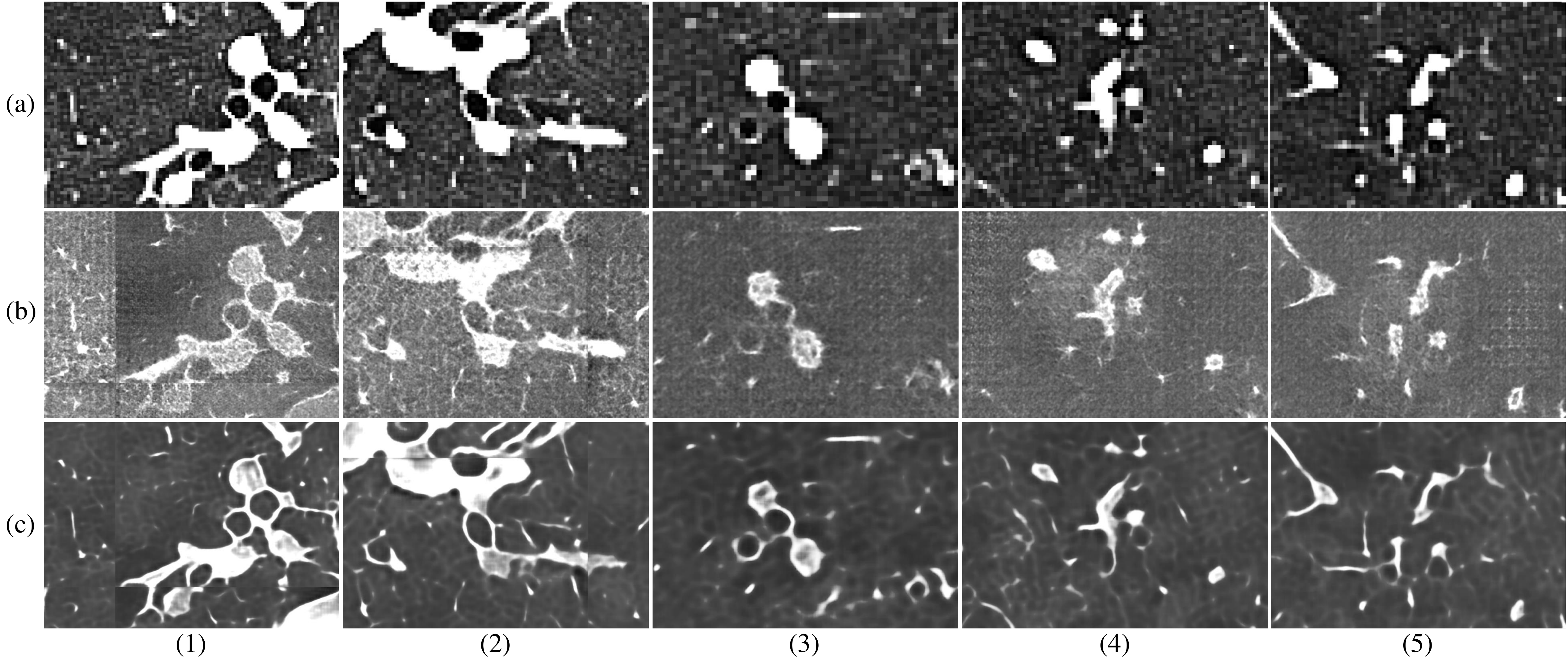}
\end{center}
\caption{More results of proposoed method and its comparison with SR-CycleGAN. Row (a): original clinical CT images; (b): SR result of conventional method (SR-CycleGAN); (c): SR result of proposed method. Columns (1), (2), (3), are images cropped from bronchus region, (4) (5) are images cropped from artery region.} \label{SRclinical_2.fig}
\end{figure}

\subsubsection{Acknowledgments}
Parts of this research is supported by ********.

\end{document}